\documentclass[aps,pre,twocolumn,twoside, amsmath,amssymb,aps,bbm,floatfix,showpacs]{revtex4-1}

\usepackage{xcolor}
\usepackage{graphicx}
\usepackage{subfigure}
\usepackage{dcolumn}
\usepackage{bm}
\usepackage{framed}
\usepackage{hyperref}
\usepackage{dsfont}
\usepackage[normalem]{ulem}
\usepackage{nicefrac}

\usepackage{nicefrac}
\usepackage{braket}
\usepackage{tikz}

\newcommand{\be}{\begin{equation}}
\newcommand{\ee}{\end{equation}}
\newcommand{\bea}{\begin{eqnarray}}
\newcommand{\eea}{\end{eqnarray}}

\begin{document}

\title{Efficiency  large deviation function  of quantum heat engines}
\author{Tobias Denzler}
\author{Eric Lutz}
 \affiliation{Institute for Theoretical Physics I, University of Stuttgart, D-70550 Stuttgart, Germany}
 
 \begin{abstract}
 The efficiency of small thermal machines is typically a fluctuating quantity. We here study the efficiency large deviation function of two exemplary  quantum heat engines, the harmonic oscillator and the two-level Otto cycles. While the efficiency statistics follows the 'universal' theory of Verley \textit{et al.} [Nature Commun. 5, 4721 (2014)] for nonadiabatic driving, we find that the latter framework does not apply in the adiabatic regime. We relate this unusual property to the perfect anticorrelation between work output and heat input that generically occurs in 
 the broad class of scale-invariant adiabatic quantum Otto heat engines and suppresses  thermal as well as quantum fluctuations.
 	 \end{abstract}
 \maketitle

Fluctuations play a central role in the thermodynamics of small systems. Contrary to macroscopic thermodynamics that describes the average behavior of a vast number of particles, microscopic  systems are characterized by stochastic variables, whose large fluctuations from mean values contain useful information on their dynamics \cite{sei12}. At equilibrium, the probability distributions of thermal observables are conveniently obtained using  the methods  of  equilibrium statistical physics  \cite{blu06}. However, their evaluation for nonequilibrium problems is often difficult. A powerful framework that allows the calculation of these distributions, both in equilibrium and nonequilibrium situations,  is provided by large deviation theory \cite{ell85,oon89,dem98,tou09}. From a physical point of view, the large deviation approach may be viewed  as a generalization of the Einstein theory of  fluctuations that relates the probability distribution to the entropy, $P(x) \sim \exp[S(x)/k]$, where $k$ is  the Boltzmann constant. On the other hand, from a mathematical standpoint, it may be regarded as an extension of the  law of large numbers and the central-limit theorem \cite{ell85,oon89,dem98,tou09}.

Large deviation techniques have found widespread application in many areas, ranging  from  Brownian motion and hydrodynamics to disordered and chaotic systems \cite{ell85,oon89,dem98,tou09}. In the past few years, they have been successfully employed to investigate the efficiency statistics of small thermal machines \cite{ver14a,ver14b,pol15,pro15,pro15a,esp15,cue15,jia15,aga15,pro15b,par15,gup17,sun19,man19,vro20}. In microscopic systems, heat, work, and, consequently, efficiency are indeed random quantities owing to the presence of thermal \cite{sei12} and, at low enough temperatures, quantum fluctuations \cite{esp09,cam11}. Understanding their fluctuating properties is therefore essential. In particular, Refs.~\cite{ver14a,ver14b} have identified 'universal' features of the  efficiency large deviation function, which exhibits  a characteristic smooth form with two extrema, including a maximum at the Carnot efficiency. The latter value is thus remarkably the least likely in the long-time limit. These predictions have been experimentally verified  for a stochastic harmonic heat engine  based on an optically trapped colloidal particle \cite{mar15}.

In this paper, we compute the efficiency large deviation function of two paradigmatic quantum thermal machines, the harmonic oscillator quantum  engine and the two-level system quantum motor \cite{kos84,gev92,scu02,scu03,lin03,kie04,rez06,qua07,dil09,aba12,zha14,wat17}. Our study is motivated by the recent experimental implementation of  a nanoscopic harmonic heat engine   using  a single trapped ion \cite{ros16} and the  realization of quantum spin-1/2 motors using NMR \cite{pet19,ass19} and trapped ion \cite{hor20} setups. We concretely consider the exemplary case of the quantum Otto cycle, a generalization of the ordinary four-stroke motor  that  has been extensively studied in the past thirty years \cite{kos17}. We find that the efficiency large deviation functions follow the 'universal' form of Refs.~\cite{ver14a,ver14b} for nonadiabatic driving. However, in the  adiabatic regime, which corresponds to maximum efficiency and may be  reached exactly for a periodically driven two-level engine or using shortcut-to-adiabaticity techniques \cite{cam14,aba17,aba17a},  we show that the large deviation functions take a markedly different shape, as the efficiency is  deterministic and equal to the macroscopic   Otto efficiency.  This result holds generically for heat engines with scale-invariant Hamiltonians that describe a broad class of single-particle, many-body and nonlinear systems \cite{gri10,jar13,cam13,def14,bea16}. We trace this unusual behavior  to the perfect anticorrelation between work output and heat input within the  engine cycle that is established  for adiabatic driving. This property completely suppresses the effects of   fluctuations. As a consequence, microscopic adiabatic quantum Otto heat engines run at the nonfluctuating macroscopic efficiency. 

\textit{Efficiency large deviation function.}  We consider a generic quantum system with a time-dependent Hamiltonian $H_t$ as the working medium of a quantum Otto engine. The  engine is alternatingly coupled to two heat baths  at inverse temperatures  $\beta_i = 1/(kT_i)$, $(i=c, h)$, where $k$ is the Boltzmann constant. The quantum Otto  cycle consists of the following four consecutive  steps \cite{kos17}: (1) Unitary expansion: the Hamiltonian is changed from $H_0$ to $H_{\tau_1}$ in a time $\tau_1$, consuming an amount of work $W_1$, (2) Hot isochore:  the system is weakly coupled to the hot  bath at inverse temperature $\beta_h$ to absorb heat $Q_2$ in a  time $\tau_2$, (3) Unitary compression: the isolated system  is driven from $H_{\tau_1}$ back to $H_0$ in a time $\tau_3$, producing  an amount of work $W_3$, and (4) Cold isochore:  the cycle  is closed by connecting  the system  to the cold bath at inverse temperature $\beta_c$,  releasing heat $Q_4$ in a time $\tau_4$.  Work and heat are positive, when  added to the system. We further assume that heating and cooling times, $\tau_{2,4}$, are longer than the relaxation time, so that the system can fully thermalize after each isochore,  as in the experimental quantum Otto engines of Refs.~\cite{pet19,ass19}.  Without loss of generality, we additionally set $\tau_1 = \tau_3=\tau$.

 The stochastic efficiency of the microscopic quantum heat engine is defined as the ratio of work output and heat input, $\eta = -W /Q_2$, where $W= W_1 + W_3$ denotes the total work \cite{ver14a,ver14b,pol15,pro15,pro15a,esp15,cue15,jia15,aga15,pro15b,par15,gup17,sun19,man19,vro20}. 
It should not be confused with the thermodynamic  efficiency of macroscopic engines, $\eta_\text{th} = -{\langle W \rangle}/{\langle Q_2 \rangle}$,  which is given by the ratio of the mean work output and the mean heat input, and is thus a deterministic quantity.
We investigate the efficiency statistics of the quantum engine in the long-time limit using large deviation theory \cite{ell85,oon89,dem98,tou09}.
Following Refs.~\cite{ver14a,ver14b}, we  write the joint distribution of  work and heat, $P_s(Q_2, W)$, as well as  the efficiency distribution $P_{s}(\eta)$, for a large number of cycles ($s\gg1$),  in the asymptotic form,  
\begin{equation}
	P_s(Q_2, W) \approx e^{-s I(Q_2,W)} \quad \text{and} \quad P_{s}(\eta) \approx e^{-s J(\eta)}.
\end{equation}
The two large deviation  functions $I(Q_2,W)$ and $J(\eta)$ describe the exponentially unlikely deviations of the variables $Q_2$, $W$ and $\eta$ from their typical values. The rate function $J(\eta)$ follows from $I(Q_2,W)$ by contraction \cite{tou09},
\begin{equation}\label{eq:jmin}
	J(\eta)=  \displaystyle{\min_{Q_2}} ~I(Q_2,-\eta Q_2).
\end{equation}
An alternative, more practical,  expression may be obtained by introducing 
the bivariate scaled cumulant generating function of the  mean heat and mean work per cycle, $q^{(s)}_2=\sum_{j=1}^s Q^{(j)}_2/s$ and $w^{(s)}=\sum_{j=1}^s W^{(j)}/s$ \cite{ver14b}:
\begin{eqnarray}\label{eq:phiqw}
	\phi(\gamma_1,\gamma_2) &=&  \displaystyle{\lim_{s \to \infty}} \frac{1}{s} \ln \langle e^{s(\gamma_1 q^{(s)}_2 + \gamma_2 w^{(s)})}\rangle \nonumber \\
						&=& \ln \langle e^{\gamma_1 Q_2 + \gamma_2 W}\rangle.
\end{eqnarray}
Using the Legendre-Fenchel transform, one then finds  \cite{ver14b},
\begin{eqnarray}\label{eq:jmin2}
	J(\eta) &=& -\displaystyle{\min_{\gamma_2}} ~ \phi(\gamma_2 \eta,\gamma_2).
\end{eqnarray}
The efficiency large deviation function  $J(\eta)$ may thus be determined from the scaled cumulant generating function $\phi(\gamma_1,\gamma_2)$. In the following, we evaluate $\phi(\gamma_1,\gamma_2)$ by taking the logarithm of the moment generating function, that is, the Wick transformed characteristic function $G(\gamma_1,\gamma_2) = \langle \exp(-i \gamma_1 Q_2 -i \gamma_2 W)\rangle$ \cite{pap91}.

\textit{Work-heat correlations.} Work output and heat input are usually correlated in a closed quantum heat engine cycle. Despite their fundamental importance, their correlations have received little attention so far \cite{camp15}. We next derive their joint probability distribution using the standard two-projective-measurement approach \cite{tal07}. In this method, energy changes of a quantum system during single realizations  of a process are identified with the difference of  energy eigenvalues obtained though projective measurements at the beginning and at the end of the process. In the quantum Otto cycle, work is performed during the unitary expansion and compression stages, while heat is exchanged during the nonunitary heating and cooling steps. We obtain the  distributions of work and heat by applying the two-projective-measurement scheme to the respective expansion, hot isochore and compression branches. The corresponding joint distribution for work output and heat input  reads accordingly \cite{sup},
\begin{eqnarray}\label{eq:p_wq}
	P(Q_2,W) &=&		\sum_{n,m,k,l} \delta \left[W - (E_m^{\tau}- E_n^0+E_l^0 - E_k^{\tau})\right] \nonumber \\
	&\times& \delta \left[Q_2 -(E_k^{\tau} - E_m^{\tau}) \right]  P_n^0(\beta_c) P_k^\tau(\beta_h) \nonumber \\
	&\times& |\bra{n} U_\text{exp}(\tau)\ket{m}|^2 |\bra{k} U_\text{com}(\tau)\ket{l}|^2 ,
\end{eqnarray}
where   $E_n^0$ ($E_k^{\tau}$) and $E_m^{\tau}$  ($E_l^0$) are the respective energy eigenvalues at the beginning and at the end of the expansion (compression) step, with corresponding unitary operator $U_\text{exp}$ ($U_\text{com}$). The thermal distribution at the beginning of the expansion (compression) stage is given by $P_n^0(\beta_c)=\exp({-\beta_c E_n^0})/Z_0$ ($P_k^\tau(\beta_h)=\exp({-\beta_h E_k^\tau})/Z_\tau$).  The  occupation probabilities $P_n^0(\beta_c)$ and $P_k^\tau(\beta_h)$ account for thermal fluctuations, while the transition probabilities $|\bra{n} U_\text{exp}(\tau)\ket{m}|^2$ and $ |\bra{k} U_\text{com}(\tau)\ket{l}|^2$ for  both quantum fluctuations and quantum dynamics \cite{jar15}.

We  study the generic features of work-heat correlations in the adiabatic regime by considering   scale-invariant Hamiltonians of the form $H_\tau= {\bm{p}^2}/{2m} + U(\bm{x},\varepsilon_\tau)$ with  $U(\bm{x},\varepsilon_\tau)=U_0 (\bm{x}/\varepsilon_\tau)/\varepsilon_\tau^2$ and scaling parameter $\varepsilon_\tau$. Such Hamiltonians describe a  large class  of single-particle, many-body and nonlinear systems with scale-invariant  spectra, $E_j^\tau= E_j^0/\varepsilon_{\tau}^2$ \cite{gri10,jar13,cam13,def14,bea16}. Taking the Fourier transform of Eq.~\eqref{eq:p_wq}, we obtain the characteristic function,
\begin{eqnarray}\label{eq:mg1g2}
	G(\gamma_1,\gamma_2)	&=& \frac{1}{Z_0 Z_\tau} \sum_n e^{\left[-\beta_c +i \varepsilon_\tau^{-2} \gamma_1 +i(1- \varepsilon_\tau^{-2}) \gamma_2\right] E_n^0} \nonumber \\
	&\times& \sum_k e^{\left[-\beta_h \varepsilon_\tau^{-2} -i \varepsilon_\tau^{-2} \gamma_1-i(1- \varepsilon_\tau^{-2}) \gamma_2\right] E_k^\tau} ,
\end{eqnarray}
with the transition probabilities $|\bra{m}U_\text{exp}\ket{n}|^2=\delta_{nm}$ and $|\bra{k} U_\text{com} \ket{l}|^2=\delta_{kl}$ for adiabatic expansion and compression. Remarkably, Eq.~\eqref{eq:mg1g2} is constant along straight lines with a slope given by  the macroscopic efficiency $\eta_\text{th}=1-\varepsilon_\tau^2$. We specifically have $G(\gamma_1,\gamma_2) = G(\gamma_1^0,\gamma_2^0)$ for $\gamma_1=\eta_\text{th}(\gamma_2-\gamma_2^0)+\gamma_1^0$. This result has profound implications for the work-heat correlations and the  large deviation properties of the quantum  engine.

\begin{figure}[t]
	\centering
	\includegraphics[width=.44\textwidth]{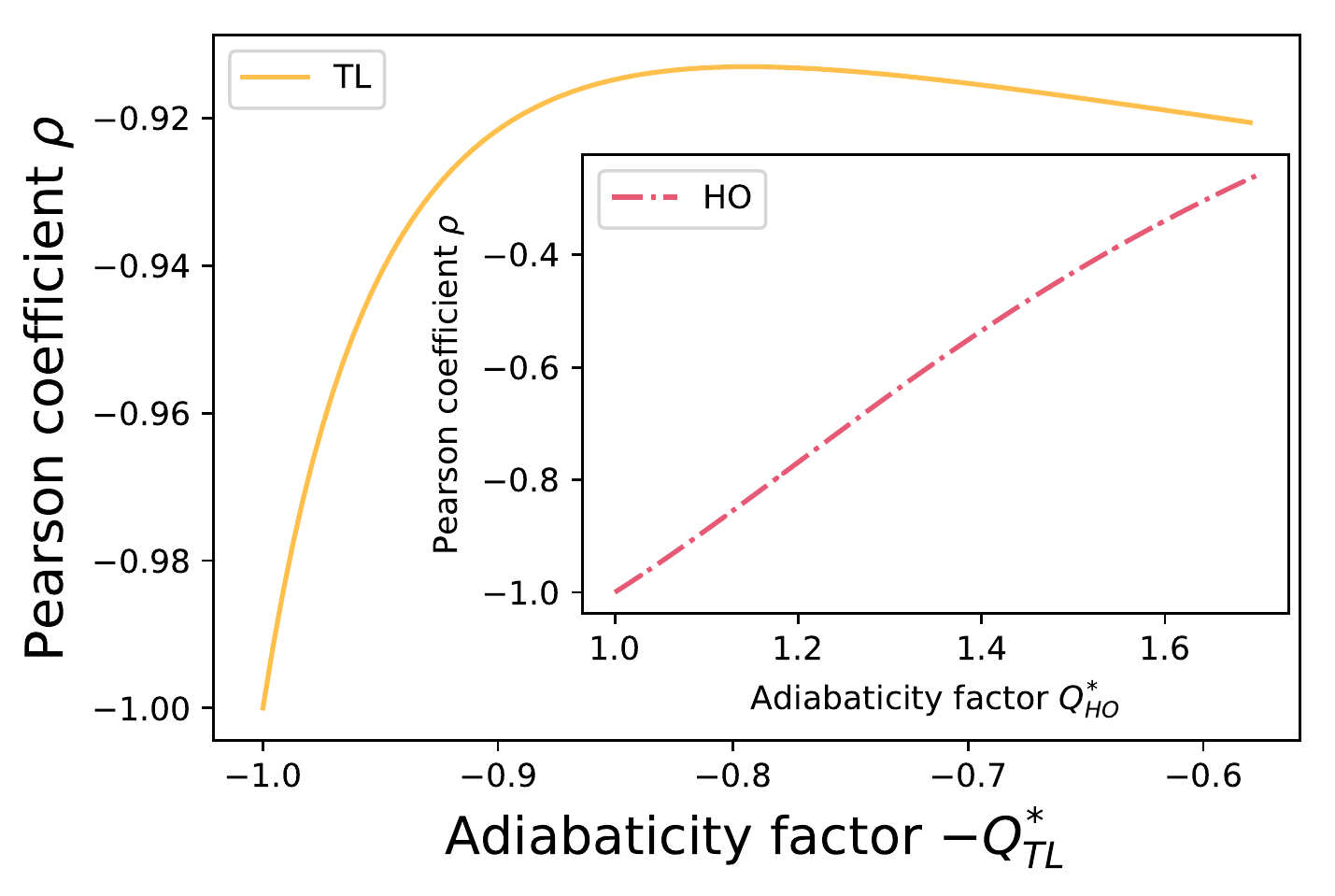}
	\caption{Work-heat Pearson coefficient  in a harmonic (red dotted-dashed) and two-level (orange solid)  Otto cycle. Work output and heat input are anticorrelated with perfect adiabatic anticorrelation, $\rho=-1$  for $Q^{*}_\text{\tiny HO}=Q^{*}_\text{\tiny TL}=1$. Parameters are  $\omega_0=\nu_0=1$, $\omega_\tau=\nu_\tau=2$, $\beta_c=3$ and $\beta_h=0.1$.}\label{fig:1}
\end{figure}

\begin{figure}[t]
	\centering
	\begin{tikzpicture}
	\node (a) [label={[label distance=-.8 cm]145: \textbf{a)}}] at (0,0) {\includegraphics[width=0.44\textwidth]{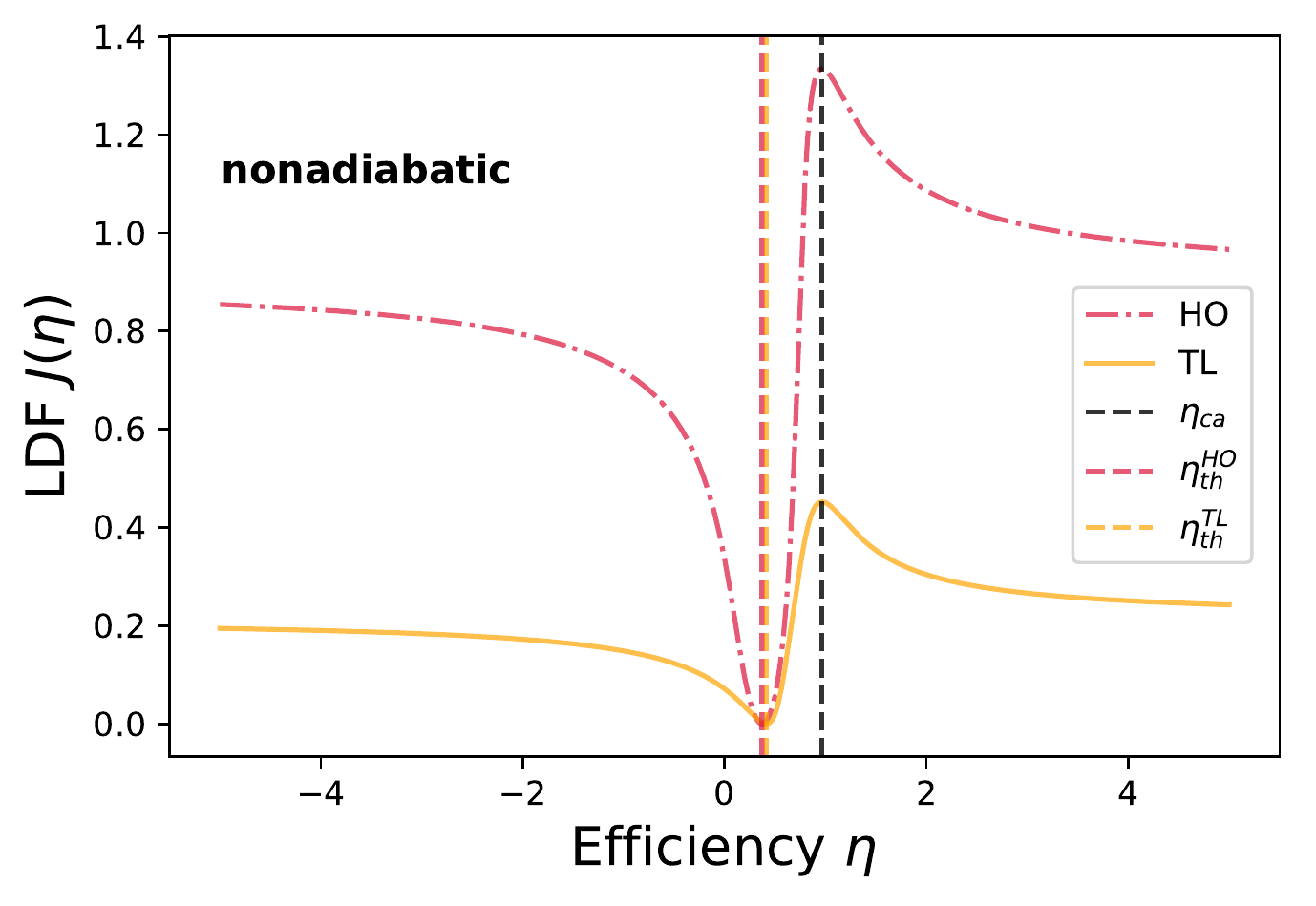}};	\node (a) [label={[label distance=-.8 cm]145: \textbf{b)}}] at (0,-5.3) {\includegraphics[width=0.44\textwidth]{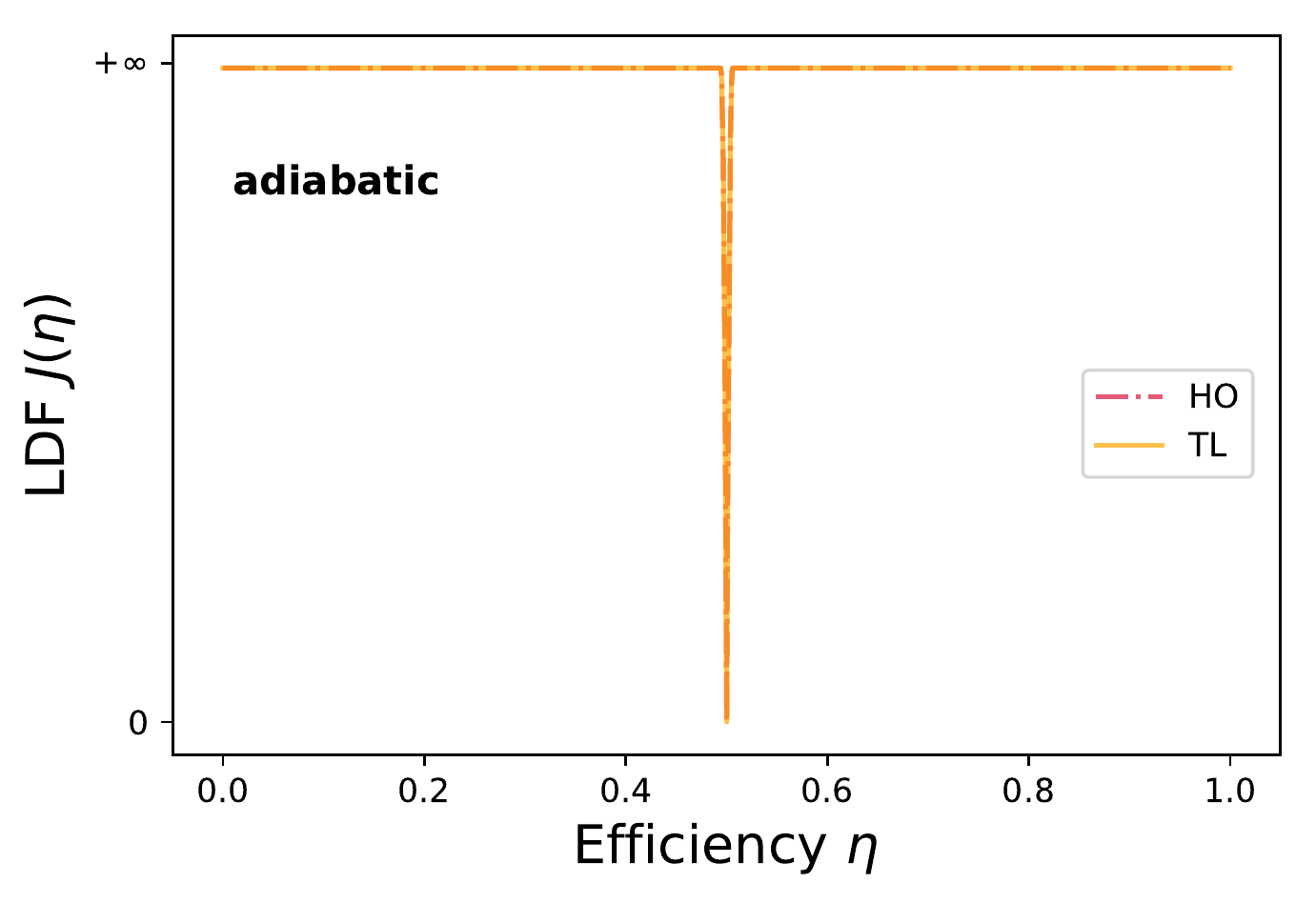}};
	\end{tikzpicture}
	\caption{Efficiency large deviation function (LDF). a) For nonadiabatic driving ($Q^{*}_\text{\tiny HO}=1.2$, $Q^{*}_\text{\tiny TL} =0.9$), $J(\eta)$ has the typical form of Refs.~\cite{ver14a,ver14b} for both   engines, with a maximum at the Carnot efficiency $\eta_\text{ca}$ and a minimum at the macroscopic efficiency $\eta_\text{th}$.  b)  For adiabatic driving ($Q^{*}_\text{\tiny HO}= Q^{*}_\text{\tiny TL} =1$),   the efficiency is  deterministic and $J(\eta)$ is infinite everywhere except  at $\eta=\eta_\text{th}$. Same parameters as in Fig.~\ref{fig:1}.} \label{fig:2}
\end{figure}

 We first remark that work output and heat input are perfectly anticorrelated in this case, with a Pearson coefficient \cite{bar89},  $\rho =\text{cov}(Q_2,W)/ \sigma_{Q_2}\sigma_{W}=-1$ (Fig.~\ref{fig:1}) \cite{sup}. On the other hand, the minimization in Eq.~\eqref{eq:jmin2} leads to a rate function that plateaus at infinity, except at the macroscopic efficiency $\eta_\text{th}$ where it vanishes (Fig.~\ref{fig:2}). As a consequence, the microscopic stochastic efficiency is deterministic and equal to the macroscopic value $\eta_\text{th}$. Adiabatic quantum Otto engines hence lie outside the universality class  of Refs.~\cite{ver14a,ver14b}. This may be  understood by noting that quantum work fluctuations are suppressed in the adiabatic regime. Thermal fluctuations are additionally canceled  by the perfect work-heat anticorrelation.

\textit{Quantum  heat engines.} Let us now examine the work-heat correlations and the efficiency large deviations, both in the adiabatic and nonadiabatic regimes, for two exactly solvable quantum Otto engines. We first evaluate the characteristic function $G(\gamma_1,\gamma_2)$  for a solvable two-level quantum motor. Inspired by the recent NMR experiments \cite{pet19,ass19}, we consider the expansion Hamiltonian, $H^\text{exp}_t= {\omega} \sigma_z/{2}+ \lambda(t) \left( \cos\omega t \,\sigma_x + \sin\omega t \,\sigma_y \right)$, that describes  a spin-$1/2$ driven by   a constant magnetic field with strength $\omega/2$ along the $z$-axis and a rotating magnetic field with varying strength $\lambda(t)$ in the ($x$-$y$)-plane,   
where $\sigma_i$, $i=(x,y,z)$, are the standard Pauli operators (with $\hbar=1$). The rotation frequency is chosen to be $\omega=\pi/2\tau$ to ensure   a complete rotation from the $x$-axis to the $y$-axis during time  $\tau$.
The amplitude of the rotating  field, $\lambda(t)= \lambda_1 \left(1-t/\tau\right)+ \lambda_2  \left(t/\tau\right)$, is increased from $\lambda_1$ at time zero to $\lambda_2$ at time $\tau$. 
This driving leads to a widening of  the energy spacing of the two-level system from 
$2\nu_0= \sqrt{4\lambda(0)^2+\omega^2}$ to $2\nu_\tau=\sqrt{4\lambda(\tau)^2+\omega^2}$. 
The compression Hamiltonian is obtained from the time reversed process, $H^\text{com}_t=-H^\text{exp}_{\tau-t}$. The  characteristic function $G(\gamma_1,\gamma_2)$ may be determined by solving the time evolution of the engine. It is explicitly given by \cite{sup},
\begin{eqnarray}\label{eq:qb_char}
	G_\text{\tiny TL}(\gamma_1,\gamma_2) &=& \frac{1}{Z_0 Z_\tau} \left \{  2 \text{cosh}(x+y) u^2 +2 \text{cosh}(x-y) v^2 \right.  \nonumber \\
	&+& 2 u v e^{-x} \text{cosh}(y) e^{-i 2 \omega_0 \gamma_2} \nonumber \\ 
	&+& 2 u v e^{x} \text{cosh}(y) e^{i 2 \omega_0 \gamma_2} \nonumber \\
	&+& u^2 e^{x-y} e^{i 2 \omega_\tau \gamma_1} e^{i 2 (\omega_0 - \omega_\tau)\gamma_2} \nonumber \\ 
	&+&v^2 e^{-x-y} e^{i 2 \omega_\tau \gamma_1} e^{-i 2 (\omega_0 + \omega_\tau)\gamma_2} \nonumber \\
	&+& 2 u v e^{-y} \text{cosh}(x) e^{i 2 \omega_\tau \gamma_1} e^{-i 2 \omega_\tau \gamma_2} \nonumber \\
	&+& u^2 e^{-x+y} e^{-i 2 \omega_\tau \gamma_1} e^{i 2 (\omega_\tau - \omega_0) \gamma_2} \nonumber \\ 
	&+&v^2 e^{x+y} e^{-i 2 \omega_\tau \gamma_1} e^{i 2 (\omega_0 + \omega_\tau) \gamma_2 }\nonumber \\
	&+& \left. 2 u v e^{y} \text{cosh}(x) e^{-i 2 \omega_\tau \gamma_1} e^{i 2 \omega_\tau \gamma_2} \right\}.
\end{eqnarray}
where $u = 1-v$ denotes the probability of no-level transition ($0\leq u \leq 1$), $x=\beta_c \nu_0$ and $y=\beta_h \nu_\tau$ \cite{sup}. The two-level engine operates  adiabatically, when the adiabaticity parameter, defined as the ratio of the nonadiabatic and adiabatic mean energies,  $Q^*_\text{\tiny TL}= 2u-1=1$ (or $u=1$). We emphasize that, since the driving is periodic,  the adiabatic regime is  here reached exactly for $\int_0^t dt' \lambda(t') = n\pi$, and not just asymptotically for large driving times \cite{sup}. Equation \eqref{eq:qb_char} contains all the information needed to investigate   the work-heat correlations and the efficiency large deviation function of the quantum two-level heat engine. 

We next consider a (unit mass) harmonic oscillator  engine with expansion Hamiltonian $H^\text{exp}_t = {p^2}/{2} +  \omega^2_t x^2/2$, where  $\omega_t$ the time-dependent  frequency that is varied from $\omega_0$ to $\omega_\tau$ in time $\tau$ according to  $\omega^2_t= \left(1 - t/\tau \right) \omega_0^2+ \omega_\tau^2  t/\tau $. The reversed compression protocol is again obtained with the replacement $t=\tau-t$. This quantum Otto engine  model is analytically solvable \cite{aba12} and we find the work-heat characteristic function   \cite{sup},
\begin{eqnarray}\label{eq:genfuncho}
&&G_\text{\tiny HO}(\gamma_1,\gamma_2) = \frac{2}{Z_0 Z_\tau}  \\
	&\times& \frac{1}{\sqrt{Q_\text{\tiny HO}^{*}(1-u_0^2)(1-v_0^2)+(1+u_0^2)(1+v_0^2)-4 u_0 v_0}} \nonumber \\
	&\times&\frac{1}{\sqrt{Q_\text{\tiny HO}^{*}(1-x_0^2)(1-y_0^2)+(1+x_0^2)(1+y_0^2)-4 x_0 y_0}}, \nonumber
\end{eqnarray}
where we have again introduced the adiabaticity parameter $Q^{*}_\text{\tiny HO}$, which is equal to 1 for adiabatic driving \cite{def08,def10}, as well as the variables $u_0 = \exp[{-\omega_0 (\beta_c+ i\gamma_2)}]$, $v_0=\exp[{i \omega_\tau (\gamma_2  - \gamma_1)}]$, 
	$x_0 = \exp[{-\beta_h \omega_\tau + i \omega_\tau (\gamma_1 - \gamma_2)}]$ and $y_0=\exp({i \gamma_2 \omega_0})$  \cite{sup}. The work-heat correlation and the efficiency large deviations of the harmonic oscillator heat engine may be examined with the help of Eq.~\eqref{eq:genfuncho}.
	\begin{figure}[t]
	\centering
	\begin{tikzpicture}
	\node (a) [label={[label distance=-.6 cm]145:\textbf{a)}}] at (0,0.2) {\includegraphics[width=0.45\textwidth]{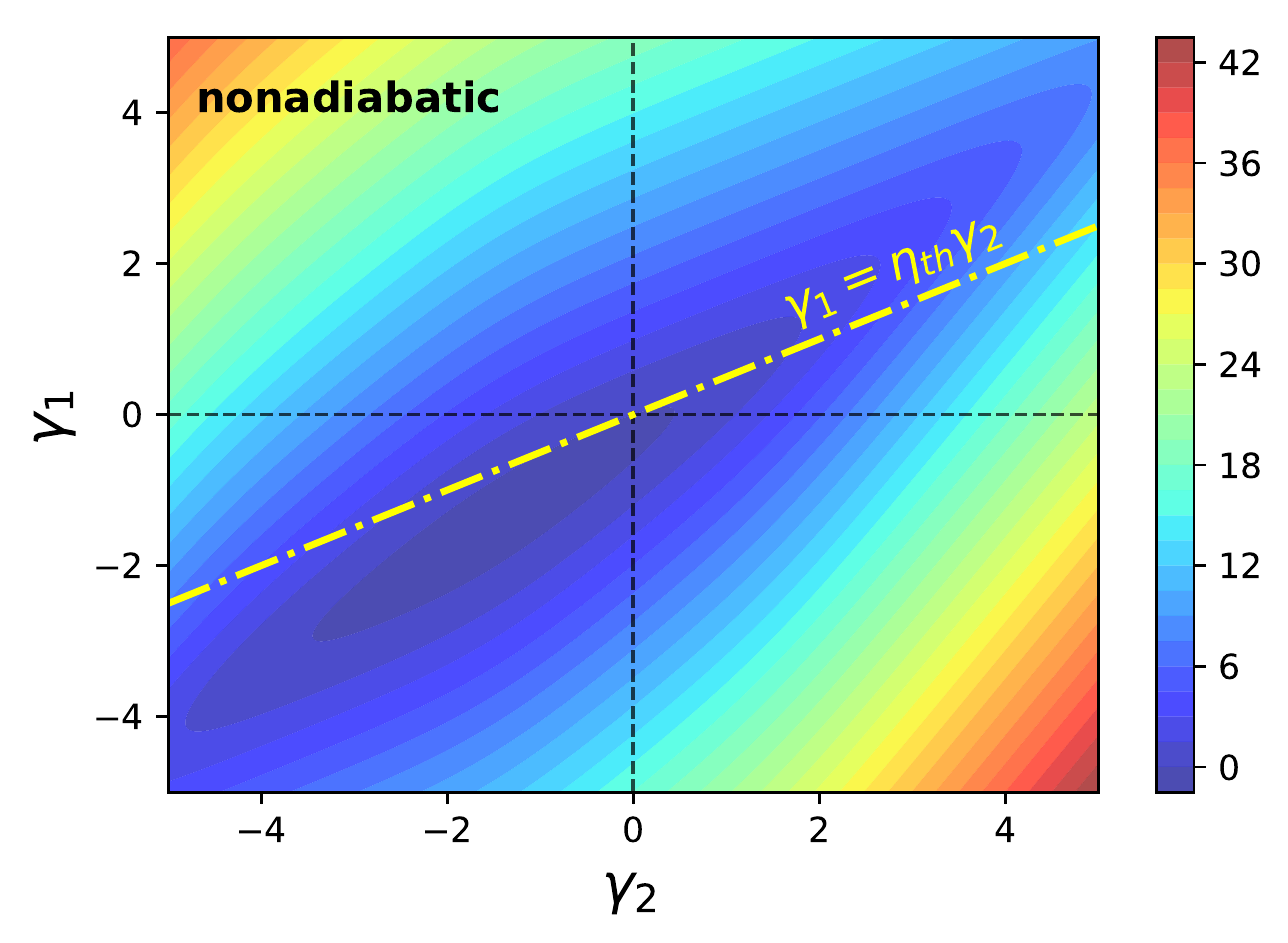}};	\node (a) [label={[label distance=-.8 cm]145:\textbf{b)}}] at (0,-5.6) {\includegraphics[width=0.475\textwidth]{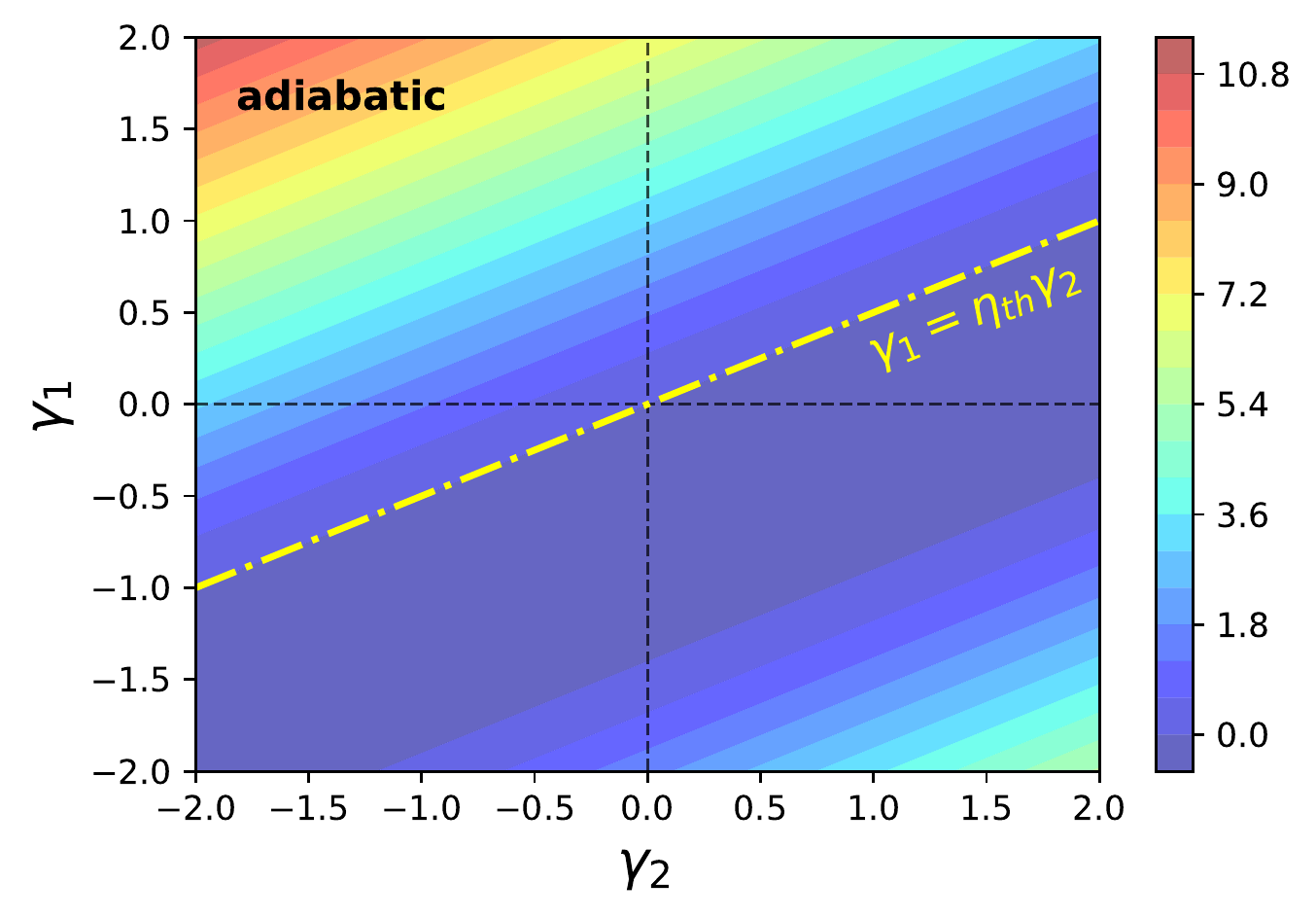}};
	\end{tikzpicture}
	\caption{Contour plot of the scaled  cumulant generating function $\phi(\gamma_1,\gamma_2)$ for the two-level  engine. a) In the nonadiabatic regime, the minimization along the line $\gamma_1 = \eta \gamma_2$  yields a unique solution, leading to the 'universal' LDF in Fig.~\ref{fig:2}a. b) In the adiabatic regime, the minimum is degenerate since the isocontours are parallel lines with slope $\eta_\text{th}$, resulting in the LDF in Fig.~\ref{fig:2}b. Same parameters as in Fig.~\ref{fig:2}.} \label{fig:3}
\end{figure}

\textit{Results.} We begin by analyzing the work-heat correlations within  the quantum Otto cycle using the Pearson coefficient. Figure \ref{fig:1} shows the correlation coefficient $\rho$ for the qubit (orange solid) and the harmonic (red dotted-dashed) quantum heat engines as a function of the respective adiabaticity parameters (we have set their frequencies equal, $\omega_j=\nu_j$, in order to compare the two cases). We observe that work output and heat input are generally negatively correlated in both examples. However, contrary to the harmonic engine, the two-level motor displays a nonmonotonous dependence on $Q^*$ due to the finite dimension of its Hilbert space. We, moreover, see that the amount of correlations increases with decreasing nonadiabaticity. In particular, work output and heat input are perfectly anticorrelated for adiabatic cycles, in agreement with the result obtained for scale-invariant engines. 

Figure \ref{fig:2} exhibits the large deviation function $J(\eta)$ for both working media. For nonadiabatic driving (Fig.~\ref{fig:2}a), we recognize the characteristic form obtained in Refs.~\cite{ver14a,ver14b}, with a maximum  at the Carnot efficiency $\eta_\text{ca}$ (the least likely value) and a minimum located at the macroscopic Otto efficiency $\eta_\text{th}$ (the most likely value). The  harmonic rate function is, furthermore, strictly above that of the qubit  (with the exception of the root at $\eta_\text{th}$),  indicating that the harmonic  heat engine converges faster towards the macroscopic efficiency $\eta_\text{th}$ than the two-level engine. By contrast, for adiabatic driving (Fig.~\ref{fig:2}b), when work output and heat input are perfectly anticorrelated, the rate function  of both systems noticeably departs from that general form: it is zero at the thermodynamic efficiency $\eta_\text{th}$ and  infinite everywhere else, confirming that  the efficiency behaves deterministically. It is important to stress that these findings are not restricted to the strict adiabatic limit \cite{sup}. They are also valid in the linear response regime, which is often used to examine the finite-time dynamics of quantum heat engines \cite{esp10,cav17,abi20}. 

A deeper understanding of the stark differences between adiabatic and nonadiabatic driving in the quantum Otto cycle may be gained by applying  the geometric approach of   Ref.~\cite{man19} to the present instance of quantum heat engines. According to Eq.~\eqref{eq:jmin2}, the rate function $J(\eta)$ is obtained for fixed $\eta$ by minimizing the cumulant generating function $\phi(\gamma_1,\gamma_2)$ along the line $\gamma_1= \eta \gamma_2$. The theory of Refs.~\cite{ver14a,ver14b} then only applies when there is a unique minimum. This is the case for nonadiabatic driving, as can be seen from the contour plot of $\phi(\gamma_1,\gamma_2)$ for the two-level quantum motor (Fig.~\ref{fig:3}a). By contrast, for adiabatic driving, the isocontours  of $\phi(\gamma_1,\gamma_2)$ are parallel lines with slope $\eta_\text{th}$ (Fig.~\ref{fig:3}b). As a result, the minimum is degenerate, leading to the plateau of the large deviation function at infinity (except at the macroscopic efficiency $\eta_\text{th}$) and the breakdown of the formalism of Refs.~\cite{ver14a,ver14b}. A similar behavior is observed for the example of the harmonic  quantum heat engine \cite{sup}.

\textit{Conclusions.} We have investigated the work-heat correlations and the efficiency statistics of the quantum Otto cycle with a working medium consisting of a two-level system or a harmonic oscillator. We have found that work output and heat input are in general negatively correlated, with perfect anticorrelation achieved for adiabatic driving. As a consequence, the microscopic quantum efficiency is equal to the deterministic macroscopic Otto efficiency and the efficiency large deviation function strongly deviates from the characteristic form obtained in Refs.~\cite{ver14a,ver14b}. These results not only hold  for quantum heat engines that operate in the adiabatic limit, such as shortcut-to-adiabaticity engines, but also in the linear response regime.  Our findings are thus important for the study of the performance of small quantum thermal machines that run close to the adiabatic regime.  

We acknowledge financial support from the German Science Foundation (DFG) under project FOR 2724.

\section*{Supplemental Material}
\section{Work-heat correlations}

We here derive the joint probability distribution of work output and heat input $P(Q_2,W)$, Eq.~(5) in the main text, using the two-projective-measurement scheme \cite{tal07}. Performing a projective energy measurement at the beginning and at end of the expansion step, we obtain the expansion work distribution $P(W_1)$,
\begin{equation}\label{eq:W1}
	P(W_1)=\sum_{n,m} \delta \left[W_1 - (E_m^\tau- E_n^0)\right] P_{n \rightarrow m}^\tau P_n^0(\beta_c),
\end{equation}
where  $E_n^0$ and $E_m^\tau$ are the respective energy eigenvalues,  $P_n^0(\beta_c)= \exp({-\beta_c E_n^0})/Z_0$ is the initial thermal occupation probability  and $P_{n \rightarrow m}^\tau= |\bra{n}U_\text{exp}\ket{m}|^2$ the transition probability between the instantaneous  eigenstates $\ket{n}$ and $\ket{m}$. The corresponding unitary  is denoted by $U_\text{exp}$.  Similarly, the probability density of the heat $Q_2$ during the following hot isochore, given the expansion work $W_1$, is equal to the conditional distribution \cite{jar04},
  \begin{equation}\label{eq:Q2}
	P(Q_2|W_1)=\sum_{k,l} \delta \left[Q_2 -(E_l^{\tau} - E_k^\tau) \right]P_{k \rightarrow l}^{\tau_2} P_k^\tau,
\end{equation}
where the occupation probability at time $\tau$ is $P_k^\tau = \delta_{km}$ when the system is in eigenstate $\ket{m}$ after the second projective energy measurement during the expansion step.  Noting that the state of the system is thermal with inverse temperature $\beta_h$ at the end of the isochore, we further have $P_{k \rightarrow l}^{\tau_2} = P_l^{\tau_2}(\beta_h)=\exp({-\beta_h E_l^{\tau}})/Z_{\tau}$.
The quantum work distribution for compression, given the expansion work $W_1$ and the heat $Q_2$, is moreover,
\begin{equation}\label{eq:W3}
	\!\!\!P(W_3|W_1,Q_2)\!=\! \sum_{i,j} \delta\! \left[ W_3 - (E_j^0 - E_i^\tau) \right] \!P_{i \rightarrow j}^\tau P_i^{\tau+\tau_2},
\end{equation}
with the  occupation probability $P_i^{\tau+\tau_2} = \delta_{il}$ when the system is in eigenstate $\ket{l}$ after the third projective energy measurement. The transition probability $P_{i \rightarrow j}^\tau=|\bra{i}U_\text{com}\ket{j}|^2$ is fully specified by the unitary time evolution operator for compression $U_\text{com}$. 

The joint  probability of having certain values of $W_3$, $Q_2$ and $W_1$ during the  cycle  follows from the chain rule for conditional probabilities, $P(W_3,Q_2,W_1) = P(W_3|Q_2,W_1)P(Q_2|W_1)P(W_1)$ \cite{pap91}. We find \cite{den20},
\begin{eqnarray}\label{eq:p_tot}
	P(W_1,Q_2,W_3)& =& \sum_{n,m,k,l} \delta \left[W_1 - (E_m^\tau- E_n^0)\right] \nonumber \\
	&\times& \delta \left[Q_2 -(E_k^\tau - E_m^\tau) \right] \delta \left[ W_3 - (E_l^0 - E_k^\tau) \right]		\nonumber \\
	&\times& |\bra{n} U_\text{exp}\ket{m}|^2 |\bra{k} U_\text{com}\ket{l}|^2 \nonumber\\
	&\times&  \frac{e^{-\beta_c E_n^0} e^{-\beta_h E_k^\tau}}{Z_0 Z_\tau}.
\end{eqnarray}
The joint distribution $P(Q_2,W)$ then follows by integrating over  the work contributions $W_1$ and $W_2$,
 $P(Q_2,W)=\int dW_1 dW_3 \delta \left[W - (W_1+W_3)\right] P(W_1,Q_2,W_3)$. 
 
 We next compute the characteristic function, Eq.~(6) of the main text, and the Pearson coefficient for adiabatic scale invariant quantum Otto heat engines with Hamiltonian $H_t= {\bm{p}^2}/{2m} + U(\bm{x},\varepsilon_\tau)$ with  $U(\bm{x},\varepsilon_\tau)=U_0 (\bm{x}/\varepsilon_\tau)/\varepsilon_\tau^2$. In the adiabatic regime, $|\bra{m}U_\text{exp}\ket{n}|^2=\delta_{nm}$ and $|\bra{k} U_\text{com} \ket{l}|^2=\delta_{kl}$,  we have,
\begin{eqnarray}
	P_\text{ad}(Q_2,W)&=& \sum_{n,k} \delta \left[W - (1-\varepsilon_\tau^{-2}) \left(E_k^0-E_n^0\right)\right] \nonumber \\
	&\times& \delta\left[Q_2 -\left(E_k^0 -E_n^0\right) \varepsilon_\tau^{-2} \right] \nonumber \\
	&\times& \frac{e^{-\beta_c E_n^0- \beta_h E_k^0/\varepsilon_\tau^2}}{Z_0 Z_\tau}.
\end{eqnarray}
The characteristic function $G(\gamma_1,\gamma_2) = \langle \exp(-i \gamma_1 Q_2 -i \gamma_2 W)\rangle$, Eq.~(6) of the main text, is readily   obtained after Fourier transformation.

The Pearson  coefficient in this case follows as,
\begin{eqnarray}\label{eq:pearson}
	\rho&=&\frac{\text{cov}(Q_2,W)}{\sigma_{Q_2}\sigma_{W}} \nonumber \\
	 &=& \frac{\langle Q_2 W\rangle -  \langle Q_2 \rangle \langle W \rangle}{\left(\langle Q_2^2 \rangle - \langle Q_2 \rangle^2\right)\left(\langle W^2\rangle - \langle W \rangle^2\right)} \nonumber \\
	&=& \frac{\left(1-\varepsilon_\tau^{-2}\right)}{|\left(1-\varepsilon_\tau^{-2}\right)|} =\pm 1.
\end{eqnarray}
We observe that work-heat correlations are always maximal in the adiabatic regime.
By further considering the heat engine conditions,
\begin{eqnarray}
	\langle Q_2 \rangle &=& \varepsilon_\tau^{-2} \sum_{n 
	\neq k} \frac{e^{-\beta_c E_n^0- \beta_h E_k^0/\varepsilon_\tau^2}}{Z_0 Z_\tau} \left(E_k^0-E_n^0\right)\geq 0  \\
	\langle W \rangle &=& \left(1-\varepsilon_\tau^{-2}\right)\sum_{n 
	\neq k} \frac{e^{-\beta_c E_n^0- \beta_h E_k^0/\varepsilon_\tau^2}}{Z_0 Z_\tau} \left(E_k^0-E_n^0\right) \leq 0,\nonumber
\end{eqnarray}
we find that $\left(1- \varepsilon_\tau^{-2} \right) \leq 0$. As a result, work output and heat input are perfectly anticorrelated for an adiabatic quantum Otto engine, $\rho=-1$. We can thus conclude that, even though the engine is still subjected to nonvanishing heat and work fluctuations, they fluctuate in unison such that its efficiency is deterministic.

\begin{figure}[t]
	\centering
	\begin{tikzpicture}
	\node (a) [label={[label distance=-1. cm]145:\textbf{a)}}]  at (0,0) {\includegraphics[width=0.44\textwidth]{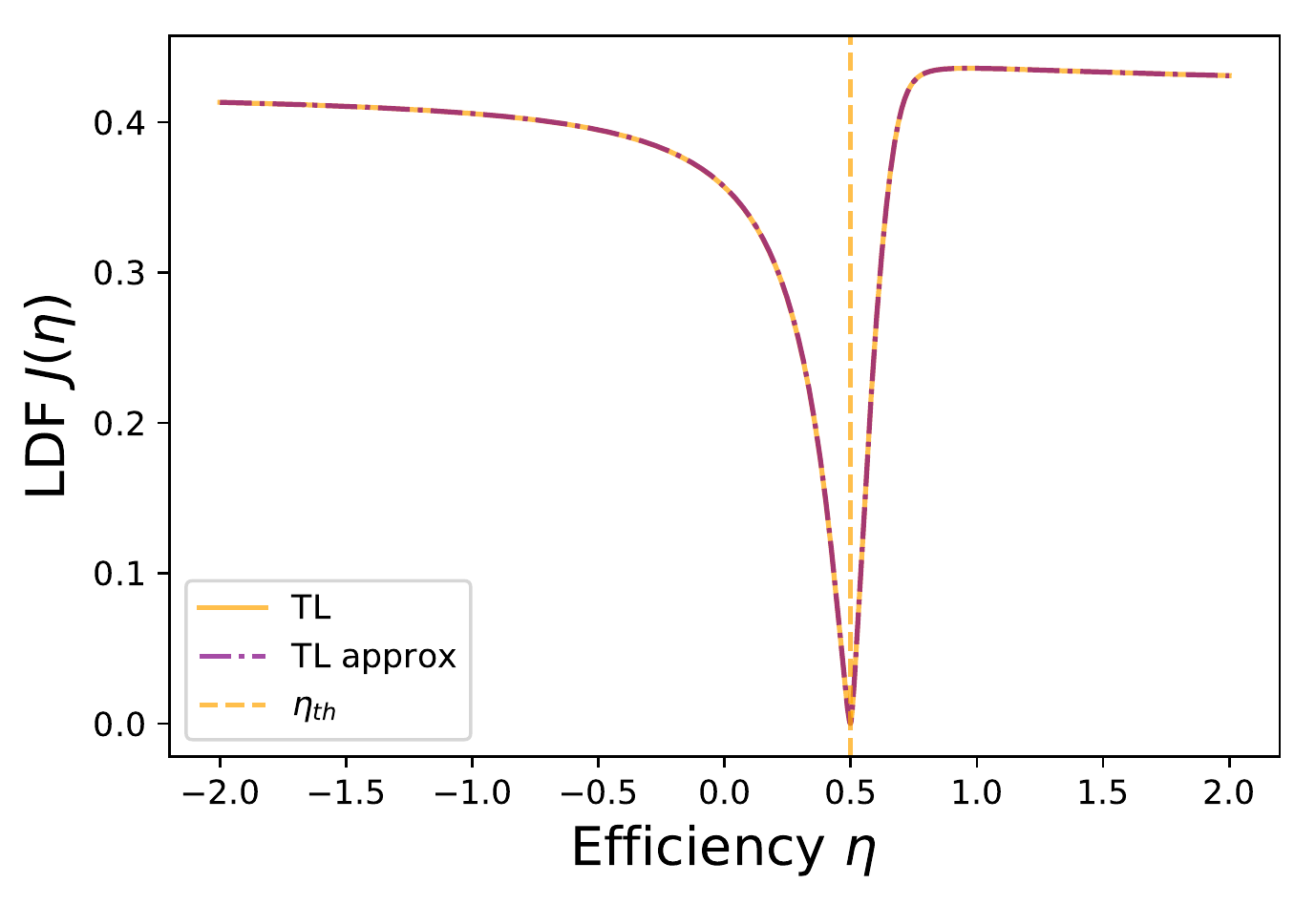}};	\node (a) [label={[label distance=-1. cm]145:\textbf{b)}}] at (0,-5.3) {\includegraphics[width=0.44\textwidth]{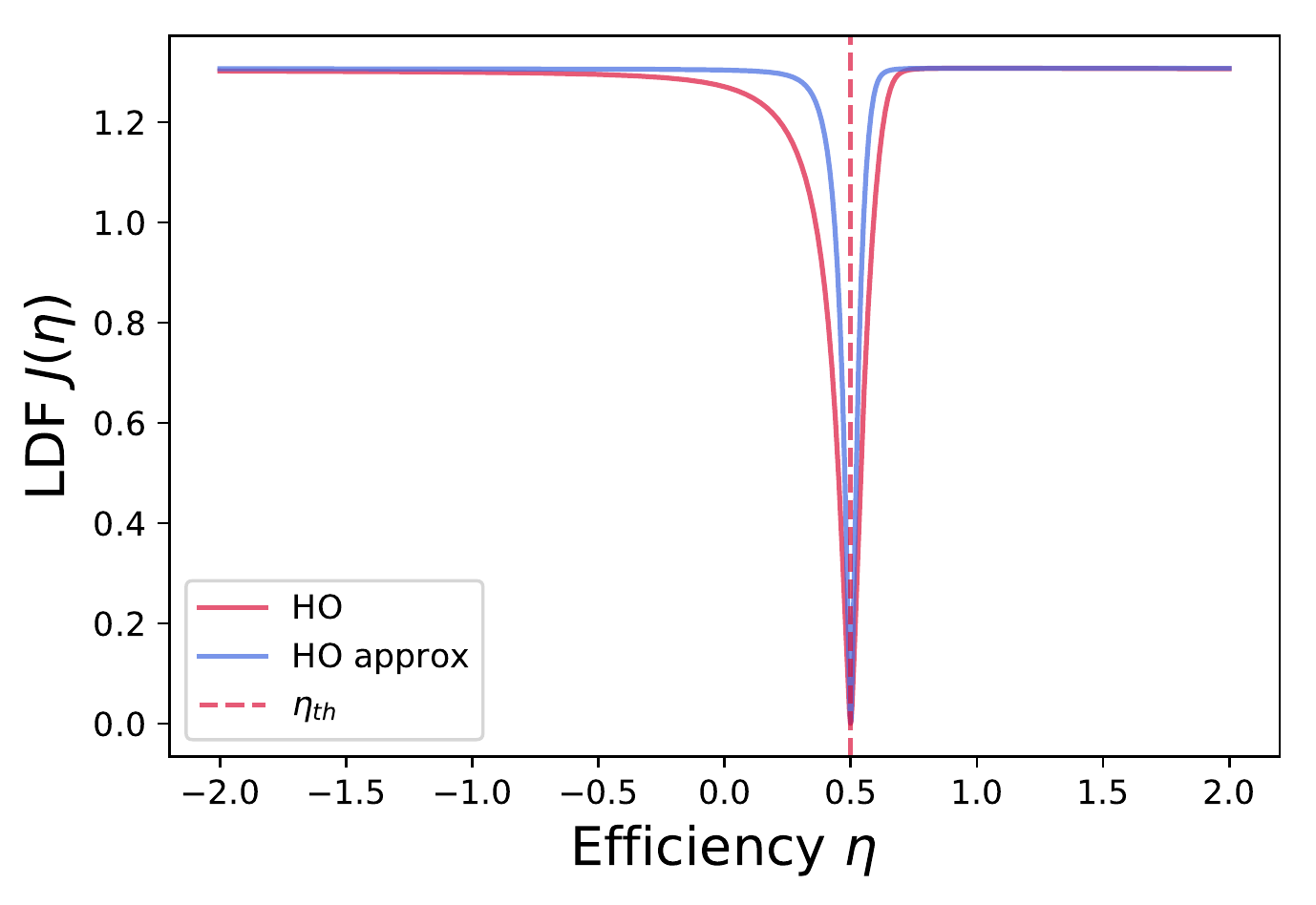}};
	\end{tikzpicture}
	\caption{Efficiency large deviation function (LDF) in the linear response regime. a) Exact $J(\eta)$ for the two-level quantum engine (orange solid) and linear approximation (purple dotted-dashed) for $Q^{*}_\text{\tiny TL}=0.998$. b) Exact $J(\eta)$ for the harmonic heat engine  (red solid) and linear approximation (blue dotted) for  $Q^{*}_\text{\tiny HO}=1.0005$. In both cases, the maximum  at the Carnot efficiency $\eta_{\text{ca}}$ effectively disappears and the peak  at the macroscopic efficiency $\eta_{\text{th}}$ is broadened. Same parameters as in Fig.~1 of the main text.} \label{fig:appfig1}
\end{figure}

\section{Characteristic functions}
We next evaluate the characteristic function $G_\text{\tiny TL}(\gamma_1,\gamma_2)$,  Eq.~(7) of the main text, for the exactly solvable two-level quantum Otto engine. The time evolution operator $U_\text{exp}$ for the expansion branch may be calculated using the methods of Refs.~\cite{den20,bar12,bar13},
\begin{equation}
\label{eq:tlsys}
	U_\text{exp} = 
	\begin{pmatrix}
		e^{-i \omega t/2}\cos I & ie^{-i \omega t/2}\sin I\\
		ie^{i \omega t/2}\sin I & e^{i \omega t/2}\cos I
    \end{pmatrix},
\end{equation}
where $I=-\int_0^t dt' \lambda(t')$ is the integral over the increasing strength of the rotating magnetic field. The operator $U_\text{com}$ follows from $U_\text{exp}$ by the replacement $t$ with $\tau -t$.  The probability of no level transition during expansion or compression  are identical  for $\tau_1=\tau_3=\tau$ and reads,
\begin{eqnarray}
	u&=& u_\text{exp}= |\bra{0}U_\text{exp}\ket{0}|=|\bra{1}U_\text{exp}\ket{1}|=\cos^2 I, \nonumber \\
	&=&u_\text{com}=|\bra{0}U_\text{com}\ket{0}|=|\bra{1}U_\text{com}\ket{1}|.
\end{eqnarray}
 The probability of a (nonadiabatic) level transition during either driving phases is accordingly $v = 1-u$. The adiabaticity parameter is defined as the ratio $Q^*_\text{\tiny TL} =\langle H_\tau \rangle_\text{nad}/\langle H_\tau \rangle_\text{ad}=2u-1$ \cite{bea16}  and is equal to $1$ for adiabatic driving, $u=1$. Inserting the above expressions for the transition probabilities into $P(Q_2,W)$ and performing the Fourier transform, $\iint dW dQ_2 ~ \exp({-i \gamma_1 Q_2 - i \gamma_2 W}) P(Q_2,W)$, then yields the characteristic function $G_\text{\tiny TL}(\gamma_1,\gamma_2)$.

The characteristic function $G_\text{\tiny HO}(\gamma_1,\gamma_2)$,  Eq.~(8) of the main text, for the exactly solvable harmonic quantum Otto engine may be directly evaluated using a result of Ref.~\cite{def10}.
The generating function of the transition probabilities for expansion is indeed given by,
 \begin{eqnarray}\label{eq:genfunc}
 	&&P(u',v')=\sum_{n,m}u_0^n v_0^m P_{n \rightarrow m}^\tau \\
 			  &=& \frac{\sqrt{2}}{\sqrt{Q^{*}_\text{\tiny HO} (1-u_0^2)(1-v_0^2)+(1+u_0^2)(1+v_0^2)-4 u_0 v_0}}.  \nonumber
 \end{eqnarray}
 and a similar expression for the compression step. We then determine the  characteristic function  $G_\text{\tiny HO}(\gamma_1,\gamma_2)$ by  comparing the terms of different powers in $(n,m,k,l)$ of the Fourier transform of Eq.~(5) of the main text with the ones in Eq.~\eqref{eq:genfunc}.

\begin{figure}[t]
	\centering
	\begin{tikzpicture}
	\node (a) [label={[label distance=-.8 cm]145:\textbf{a)}}]  at (0.2,0) {\includegraphics[width=0.45\textwidth]{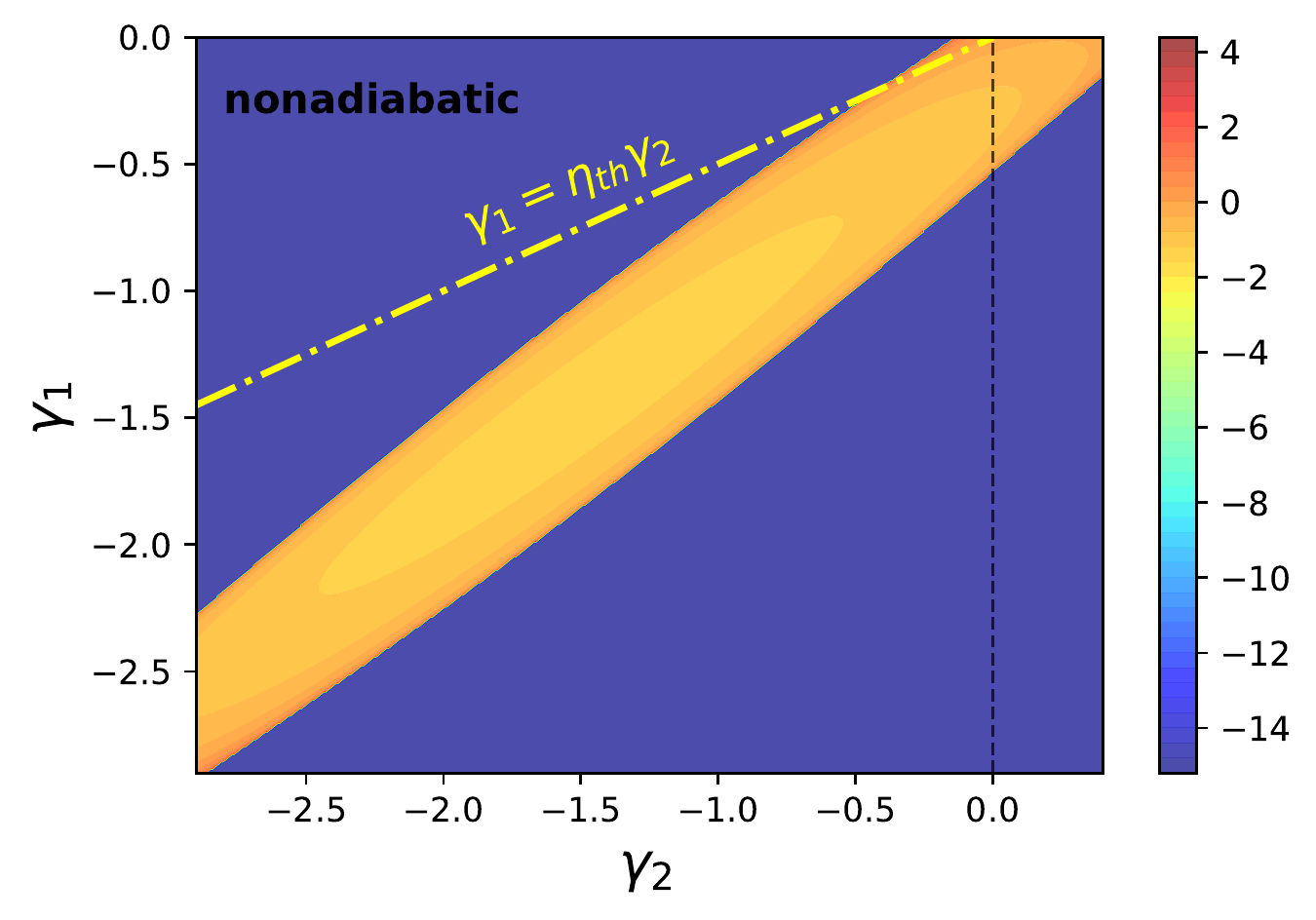}};	\node (a) [label={[label distance=-.55 cm]145:\textbf{b)}}]  at (.4,-5.5) {\includegraphics[width=0.45\textwidth]{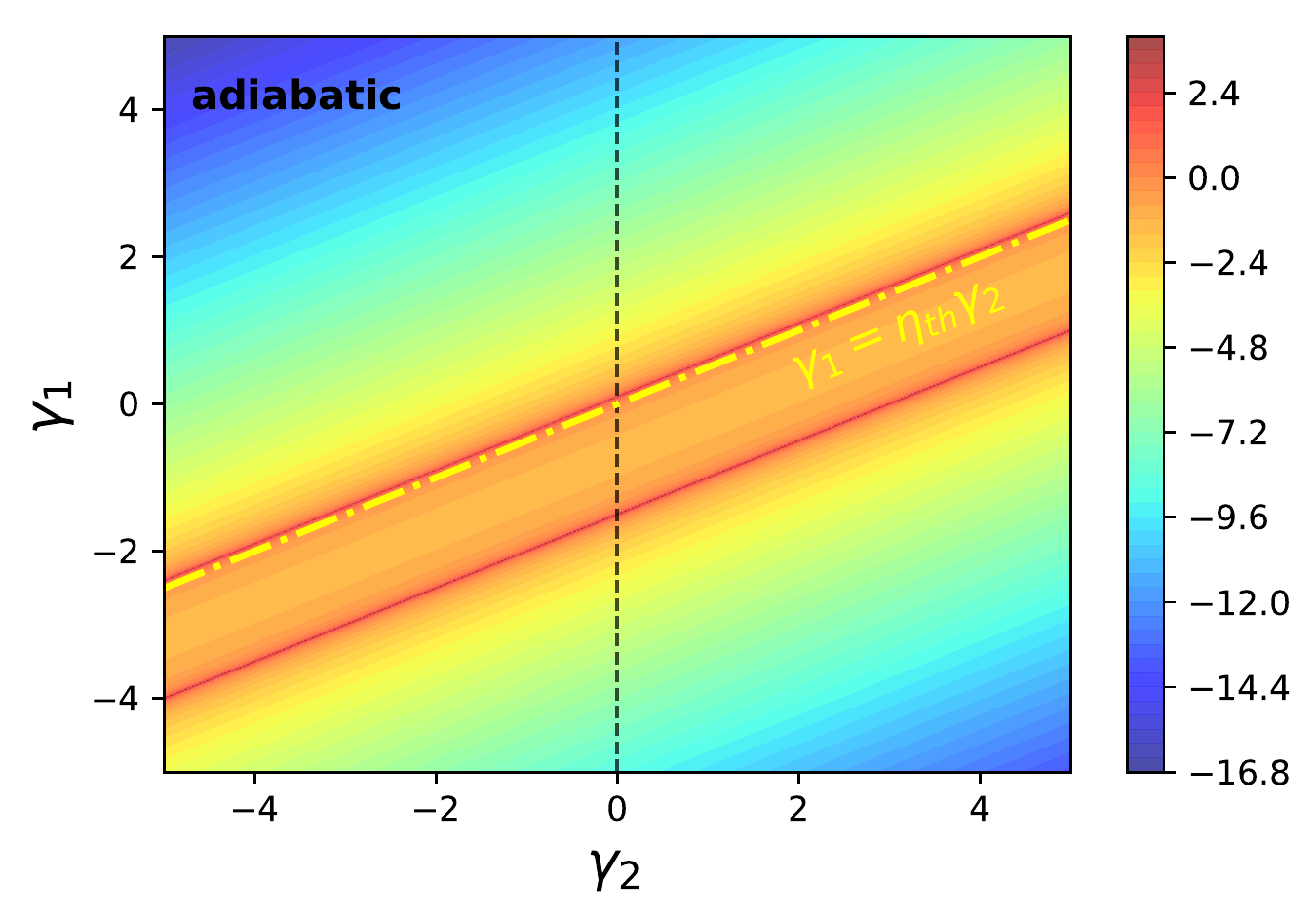}};
	\end{tikzpicture}
	\caption{Contour plot of the scaled  cumulant generating function $\phi(\gamma_1,\gamma_2)$ for the harmonic quantum engine. a) In the nonadiabatic regime ($Q^{*}_\text{\tiny HO}=1.2$), the minimization along the line $\gamma_1 = \eta \gamma_2$  yields a unique solution. b) In the adiabatic regime ($Q^{*}_\text{\tiny HO}=1$), the minimum is degenerate since the isocontours are parallel lines with slope $\eta_\text{th}$.}\label{fig:appfig2}
\end{figure}

\section{Linear-response regime}
The performance of (nonadiabatic) finite-time quantum heat engines is often analyzed in the linear response regime, that is, in a first-order expansion around the adiabatic limit \cite{esp10,cav17,abi20}. We show in this section that the large deviation function $J(\eta)$ still deviates from the general form of Refs.~\cite{ver14a,ver14b}. For the two-level Otto engine, a Taylor expansion in first-order around $u=1$ yields the work-heat characteristic function,
\begin{eqnarray}
	G_\text{\tiny TL}^\text{lin}(\gamma_1,\gamma_2) &=& \frac{1}{Z_0 Z_\tau}  \left\lbrace e^{2 i \gamma _2 \left(\omega _0-\omega _{\tau }\right)+2 i \gamma _1 \omega _{\tau }+x-y} \right. \nonumber \\
	&+& e^{-2 i \gamma _1 \omega _{\tau }+2 i \gamma _2 \left(\omega _{\tau }-\omega _0\right)-x+y}+2 \cosh (x+y) \nonumber \\
	&+& 2 (u-1)\left[ e^{2 i \gamma _2 \left(\omega _0-\omega _{\tau }\right)+2 i \gamma _1 \omega _{\tau }+x-y} \nonumber \right. \\
	&+& e^{-2 i \gamma _1 \omega _{\tau }+2 i \gamma _2 \left(\omega _{\tau }-\omega _0\right)-x+y} \nonumber \\
	&-& \cosh (x) e^{2 i \gamma _1 \omega _{\tau }-2 i \gamma _2 \omega _{\tau }-y} \nonumber \\
	&-&\cosh (x) e^{-2 i \gamma _1 \omega _{\tau }+2 i \gamma _2 \omega _{\tau }+y} \nonumber \\
	&-& \cosh (y) e^{-x-2 i \gamma _2 \omega _0} - \cosh (y) e^{x+2 i \gamma _2 \omega _0} \nonumber \\
	&+& \left. \left. 4 \cosh (x+y) \right] \right\rbrace.
\end{eqnarray}
where the parameters $x, y, u, v$ are unchanged.

On the other hand, a Taylor expansion in first-order around $Q^{*}_\text{\tiny HO}=1$ yields the work-heat characteristic function for the harmonic quantum Otto heat engine,
\begin{eqnarray}
	G_\text{\tiny HO}^\text{lin}(\gamma_1,\gamma_2) &=& \frac{1}{Z_0 Z_\tau}\left\lbrace\frac{1}{\left(1-u_0 v_0\right) \left(1-x_0 y_0\right)}  \right.\nonumber \\
	&+&  \frac{1-q}{4 \left(1-u_0 v_0\right){}^3 \left(1-x_0 y_0\right){}^3} \nonumber \\
			&\times& \left[\left(1-x_0^2\right) \left(1-y_0^2\right) \left(1-u_0 v_0\right){}^2 \right. \nonumber \\
			&+& \left. \left. \left(1-u_0^2\right) \left(1-v_0^2\right) \left(1-x_0 y_0\right){}^2\right]  \right\rbrace,
\end{eqnarray}
where the parameters $x_0, y_0, u_0, v_0$ are also unchanged.

The corresponding approximate and exact large deviation functions $J(\eta)$ are shown in Fig.~\ref{fig:appfig1}a) for the two-level engine and in Fig.~\ref{fig:appfig1}b) for the harmonic motor. We observe  in both cases that the maximum  at the Carnot efficiency $\eta_{\text{ca}}$ has effectively disappeared and that the narrow  peak at the minimum located at the macroscopic efficiency $\eta_{\text{th}}$ has instead broadened.

\section{Harmonic scaled cumulant generating function}

We finally show the contour plots of the scaled cumulant generating function $\phi(\gamma_1,\gamma_2)$ of the harmonic  quantum Otto engine in the nonadiabatic (Fig.~{\ref{fig:appfig2}a)  and adiabatic (Fig.~{\ref{fig:appfig2}b) regimes.  They are qualitatively similar to those of the two-level quantum motor represented in Fig.~3 of the main text. In the nonadiabatic case, we find regions in the $(\gamma_1,\gamma_2$)-plane for which the cumulant generating function is undefined (dark blue), contrary to what happens for the two-level Otto engine. This might lead to additional deviations from the 'universal' theory of Refs.~\cite{ver14a,ver14b} as those already pointed out in Ref.~\cite{man19}. In the adiabatic case, we again observe   parallel lines with slope $\eta_{\text{th}}$, leading to a degenerate minimum in the minimization procedure of the rate function $J(\eta)$.

 \end{document}